\documentclass[preprint]{aastex}

\shorttitle{Gravitational lensing of S-stars} \shortauthors{Bozza
\& Mancini}

\begin{document}

\title{Gravitational lensing of stars orbiting \\
the Massive Black Hole in the Galactic Center}

\author{V. Bozza\altaffilmark{1} and L. Mancini\altaffilmark{2}}
\affil{Dipartimento di Fisica ``E.R. Caianiello'', Universit\'a di
Salerno, Italy. \\
Istituto Nazionale di Fisica Nucleare, Sezione di Napoli, Italy. }

\altaffiltext{1}{valboz@physics.unisa.it}
\altaffiltext{2}{lmancini@physics.unisa.it}

\begin{abstract}
The existence of a massive black hole in the center of the Milky
Way, coinciding with the radio source Sgr A*, is being established
on more and more solid ground. In principle, this black hole,
acting as a gravitational lens, is able to bend the light emitted
by stars moving within its neighborhood, eventually generating
secondary images. Extending a previous analysis of the
gravitational lensing phenomenology to a new set of 28 stars,
whose orbits have been well determined by recent observations, we
have calculated all the properties of their secondary images,
including time and magnitude of their luminosity peaks and their
angular distances from the central black hole. The best lensing
candidate is represented by the star S6, since the magnitude of
its secondary image at the peak reaches $K=20.8$, with an angular
separation of 0.3 mas from the central black hole, that is just at
the borders of the resolution limit in the $K$ band of incoming
astronomical instruments.
\end{abstract}

\keywords{Gravitational lensing --- Black hole physics --- Stars:
individual (\objectname{S1}, \objectname{S2}, \objectname{S4},
\objectname{S5}, \objectname{S6}, \objectname{S8},
\objectname{S9}, \objectname{S12}, \objectname{S13},
\objectname{S14}, \objectname{S17}, \objectname{S18},
\objectname{S19}, \objectname{S21}, \objectname{S24},
\objectname{S27}, \objectname{S29}, \objectname{S31},
\objectname{S33}, \objectname{S38}, \objectname{S66},
\objectname{S67}, \objectname{S71}, \objectname{S83},
\objectname{S87}, \objectname{S96}, \objectname{S97})
--- Galaxy: center}

\section{Introduction}
One of the most fascinating arenas for testing General Relativity
in the neighborhood of a real astrophysical black hole is
undoubtedly represented by Sgr A*, the massive black hole (MBH) at
the center of our Galaxy, whose existence has been now strongly
proven by the combination of radio, sub-mm, X-ray, and Near
Infrared (NIR) observations. While in the radio domain Sgr A*
appears as a steady and motionless compact source \citep{Reid07},
its X-ray, sub-mm, and NIR counterparts are variable since they
exhibit outbursts of energy ({\it flares}), which typically last
for a few hours and occur several times a day
\citep{Baganoff01,Baganoff03,Genzel03,
Ghez04,Clenet05,Eckart06,Eckart08,Hamaus08,Marrone08,Yusef08}.
These flares are likely due to energetic events arising very close
to the central MBH, on a scale of a few Schwarzschild radii, and
they can be interpreted as due to emission from matter in
relativistic orbits around Sgr A*. Moreover, by using large class
telescopes, such as the ESO Very large Telescope (VLT) and the
Keck telescopes, and thanks to the adaptive optics, it has been
possible to observe a multitude of stars in the $K$ band (the
so-called S-stars) moving in the gravitational potential of the
Galactic MBH \citep{Schodel03,Ghez05,Eisenhauer05}. The entire
system is perfectly described by a single point mass and Newton
gravity. Very recently, \citet{Gillessen08} have been able to
determine the orbits of 28 S-stars, resting upon the results of 16
years of monitoring of their motions. In particular, it has been
possible to observe the whole orbit of the famous star S2, since
it has completed a full revolution around Sgr A*. The value of the
Galactic MBH mass, extrapolated from these improved data, is
$M=4.31\pm0.36 \times 10^{6}$ M$_{\sun}$.

Since the theory of General Relativity predicts that the central
MBH can act as a powerful gravitational lens on S-stars, some of
them, depending on the knowledge of their orbital parameters, have
been the object of careful investigations. The strategy of this
study is very clear. In fact, knowing the position of the S-stars
in space as a function of time, it is easy to predict the time of
their periapse, that is the closest approach to Sgr A*, and look
for possible alignments between one of them and the lens,
corresponding to the MBH, along the line of sight of an observer
on the Earth. In this way, it is possible to foresee at any time
where to look for a secondary image, what its apparent magnitude
should be, and, thanks to our knowledge of their orbits, easily
predict the best time to observe them.

S2 was the first star to be studied as a possible source for
gravitational effects due to the central MBH
\citep{DePaolis03,Bozza04}. The light curve for its secondary
image, as well as the first two relativistic images, were
calculated in the Schwarzschild black hole hypothesis. The
analysis was then enlarged by \citet{Bozza05}, who achieved the
light curves for the secondary images of five more S-stars. An
important fact that emerges from the latter study is that the
light curves are peaked around the periapse epoch, but two
subpeaks may arise in nearly edge-on orbits, when the source is
behind Sgr A* ({\it standard lensing}) or in front of it ({\it
retrolensing}). In this respect, the star S14 represents an
emblematic case.

In the present paper we want to extend the previous analysis of
\citet{Bozza05} to the new set of 28 S-stars, provided by
\citet{Gillessen08}. We will see that, within these stars, S6
turns out to be a particularly intriguing source for gravitational
lensing, since its secondary image is potentially observable by
incoming interferometric instruments.

\section{Lens Equation and Position of the Images} \label{Sec
Lens}

Our aim is to calculate the position and the luminosity of the
secondary images of the 28 S-stars as functions of time, generated
by the MBH at the center of our Galaxy. The secondary image
appears on the opposite side of the black hole with respect to the
source and is thus formed by light rays passing behind the MBH.
Defining the deflection angle $\alpha$ as the angle between the
asymptotic directions of motion of the light rays, before and
after the rendezvous with the black hole, the lens equation is
\citep{Oha,Bozza04}
\begin{equation}
\gamma=\alpha(\theta)- \theta
-\arcsin\left(\frac{D_{\mathrm{OL}}}{D_{\mathrm{LS}}}\sin\theta\right),
\label{Lens Eq}
\end{equation}
where $\gamma$ is the alignment angle of the source, that is the
angle between the line connecting the source to the lens and the
optical axis, defined as the line connecting the observer to the
lens. The geometry of the lensing configuration is such that
$\gamma$ runs from 0 (perfect alignment) to $\pi$ (perfect
anti-alignment). $\theta$ is the impact angle for the light rays
as seen from the observer, $D_{\mathrm{OL}}$ and $D_{\mathrm{LS}}$
are the distances between the observer and the lens, and the lens
and the source, respectively. $D_{\mathrm{OL}}$ is given by the
distance of the Sun to the Galactic center, which is currently
estimated to be $R_{0}=8.33\pm0.35$ kpc \citep{Gillessen08}.
Following \citet{Bozza05}, we obtain $D_{\mathrm{LS}}$ and
$\gamma$ as functions of time by an accurate modelling of the
Keplerian orbit of the S-stars around the MBH, taking into account
the measured values of the orbital parameters. For all the
details, the reader is referred to \citet{Bozza05}.

As explained in \citet{Bozza08}, the lens equation (\ref{Lens Eq})
is the closest to the exact treatment of deflection by a
spherically symmetric gravitational field. The relative error in
the position of the images committed by using Eq. (\ref{Lens Eq})
is at most of the order of $(R_\mathrm{g}/D_{\mathrm{LS}})^2$,
where $R_\mathrm{g}=2GM/c^2$ is the gravitational radius of the
black hole. This number stays lower than $10^{-6}$ in the worst
cases.

The deflection angle is calculated assuming that the black hole is
spherically symmetric, thus being described by a Schwarzschild
metric. Since the secondary image of an S-star demands a
deflection angle in the whole range $[0,\pi]$, we cannot adopt the
weak-field or the strong deflection limit approximation, but we
have to calculate the exact deflection angle in terms of elliptic
integrals \citep{Darwin59}. Finally, the magnification of a
secondary image is given by the general formula
\citep{Oha,Bozza04}
\begin{equation}
\mu=\frac{D_{\mathrm{OS}}^2}{D_{\mathrm{LS}}^2}\frac{\sin\theta}{
\frac{d \gamma}{d\theta} \sin{\gamma}}, \label{mu}
\end{equation}
where $D_{\mathrm{OS}}$ is the distance between the observer and
the source.

%
\section{Results}
In the NIR band, the Sgr A* stellar cluster has been originally
observed with the ESO's NTT starting from 1992, then with the Keck
from 1995, and with the VLT from 2002. The good synergy between
the adaptive optics and the 8m/10m-class telescopes
\citep{Ghez01,Schodel02} allowed to push the astronomical imaging
and the spectroscopy for point sources to a deep sensitivity, that
is $K\sim16-17$ at an angular resolution of $50-150$ mas
\citep{Eckart02}. Thanks to these powerful instruments of
astronomical investigation, it was possible to monitor the S-stars
for more than 16 years and reconstruct for some of them their
Keplerian orbits around the Galactic MBH \citep{Gillessen08}. The
orbital parameters of these 28 stars are summarized in
table~\ref{tbl-1}.

Following the steps described in Section \ref{Sec Lens}, we have
calculated the light curves of the secondary images of the
mentioned 28 stars. The main features of these images are
presented in table~\ref{tbl-2}, allowing an immediate reading of
the results. Besides the values of the minimum magnitude for the
secondary images and the epochs at which they peak, we also report
the time spent by the secondary images when their luminosity is
greater than half of the maximum, the alignment angle and the
distance from the MBH of the source stars at the peak, and the
maximal angular distance of the secondary images from the apparent
shadow of the black hole (whose angular radius is
$3\sqrt{3}R_\mathrm{g}/2D_\mathrm{OL}$ \citealt{Darwin59}).

\begin{figure}
\centering{\includegraphics{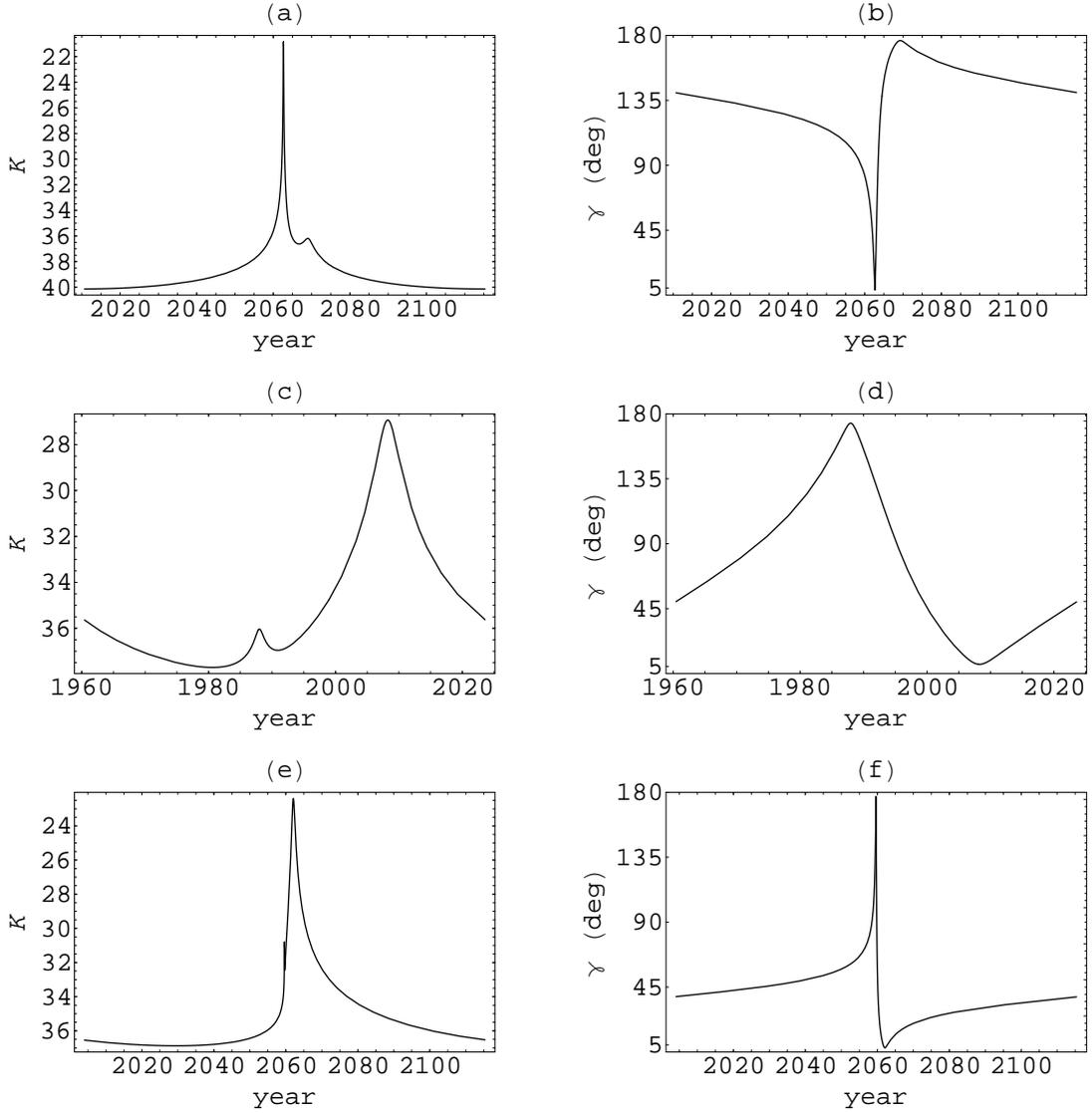}}
 \caption{Left column: light curves for the secondary images
 of S6 (a), S17 (c), S27(e). Right column: the alignment angle
 $\gamma$ as a function of time for
 S6 (b), S17 (d), S27(f).}
 \label{Fig 1}
\end{figure}

Among all the S-stars analyzed, we concentrate our attention on
three particular cases\footnote{The case of S14 was already
examined in \cite{Bozza05}. With respect to the estimates
presented there, there is no significant change with the new
data.}: S6, S17, and S27, characterized by a value of the orbit
inclination very close to $90^{\circ}$. This feature qualifies
them as very appealing candidates for gravitational lensing by the
MBH. In Figure~\ref{Fig 1} we report the light curves for the
secondary images of the three mentioned stars (left column) and
the alignment angles as functions of time (right column).

S6 -- This star is indeed the most interesting one, since its
orbit is nearly edge-on ($i=86.44^\circ$), and its eccentricity
brings it quite close to the black hole (up to 4800
$R_\mathrm{g}$). The best alignment and anti-alignment times are
very close to the periapse epoch and are responsible for two
subpeaks, see Figure~\ref{Fig 1}a. By examining the
Figure~\ref{Fig 1}b, we notice that first we have the alignment
peak ($\gamma \sim 0$), and then the anti-alignment peak ($\gamma
\sim \pi$). What makes S6 particularly interesting is the high
brightness attained by the secondary image. At the best alignment
time, that will occur in 2062, the secondary image has a magnitude
$K=20.8$, and the maximal angular distance from the apparent
shadow of the MBH is about $0.316$ mas, according to the current
estimates of the orbital parameters.

S17 -- The orbit of this star is nearly edge-on as well
($i=96.44^\circ$). Yet, contrarily to the previous case, the
eccentricity of S17 is very low and the periapse occurs in
proximity of one of the nodes. For this reason, the alignment and
anti-alignment peaks occur at opposite epochs in the period and
neither of them is particularly bright (Figures~\ref{Fig 1}c and
\ref{Fig 1}d).

S27 -- From a gravitational-lensing point of view, the inclination
of the orbit of this star is the most favorable among all the
known S-stars ($i=92.91^\circ$). Moreover, the eccentricity of the
orbit is very close to one. However, the periapse occurs in
correspondence to the anti-alignment epoch (compare
Figure~\ref{Fig 1}e with Figure~\ref{Fig 1}f). By reflex, the
anti-alignment peak is very tight and hardly visible in
Figure~\ref{Fig 1}e because S27 moves very fast at that time.
Conversely, the alignment peak is not so bright ($K=22.4$) as in
the case of S6, but anyway better than S14 and S17. It is
interesting to note that the angular separation of the secondary
image of S27 reaches almost 0.4 mas at the best alignment time,
which puts this star in a very good position for possible
observations.

%
\section{Discussion}
Thanks to the great technological advances earned by the modern
big-class telescopes, it has been possible to have access to very
remote zones of the Milky Way, like the neighborhood of its
center, where the radio source Sgr A* lies. The discovery that
most of the detected stars in this region move along tight
Keplerian orbits is a strong indication of the existence of a very
compact object of more than $4 \times 10^{6}$ M$_{\sun}$,
currently recognized as a massive black hole, in the geometrical
center of the Milky Way. This great disclosure opens an amazing
way of escape from the meanders of theoretical speculations and
allows us to study gravitational lensing by a black hole on a real
case and test General Relativity in its strong field regime. The
idea of using the star S2 as a potential source to be lensed by
the central MBH, was conceived originally by \citet{DePaolis03},
and then acquired and rigorously developed by
\citet{Bozza04,Bozza05}, who considered other S-stars as further
possible sources for this phenomenology. In this paper, we extend
the previous analysis to a new set of 28 S-stars provided by
\citet{Gillessen08}. Whereas most of these stars have undetectable
secondary images, there are some cases in which the lensed images
have a significant probability of being detected by incoming
instruments, as can be deduced from table~\ref{tbl-2}. Among
these, S6 represents the best case since its secondary image at
the peak reaches $K=20.8$, with an angular separation of 0.3 mas
from the central black hole. Actually, the perspectives for
observing such an event are not so far from what the modern
technology is developing. In fact, in 2013 the new
second-generation VLTI instrument, called GRAVITY, will be
scientifically operative at the ESO Paranal observational site
\citep{Eisenhauer08}. This instrument, specifically designed to
observe highly relativistic motions of matter close to the event
horizon of Sgr A*, will interferometrically combine the NIR light
collected by the four-units telescope of the VLT, exploiting all
the potential of this extremely powerful ESO observatory. The
resolution that will be obtained in the NIR band is $\sim 3$ mas
for objects that can be as faint as $17<K<19$. Moreover, in its
astrometric mode, GRAVITY will have an accuracy of $10$
$\mathrm{\mu as}$, allowing to study the motions of celestial
objects to within a few times the event horizon size of the
Galactic MBH. There are still many years before the secondary
images of the best known S-stars will shine again (2047 for S14,
2062 for S6 and S27). Surely, new S-stars will be discovered and
followed along their orbits, eventually providing even better
candidates for gravitational lensing. In the meanwhile, we have no
doubt that the astronomical community will fortify its stock of
observational facilities with new advanced instruments, which will
be able to catch this extraordinary kind of gravitational lensing
events.

\begin{table*}
\centering
\begin{scriptsize}
\begin{tabular}{lcccccccc}
\tableline %
\tableline %
Star & $a['']$ & $e$ & $i[^\circ]$ &
$\Omega[^\circ]$ & $\omega[^\circ]$ & \
$t_{\mathrm{P}}[\mathrm{yr}]$ & $T[\mathrm{yr}]$ & $K$ \\
\tableline %
S1 & 0.508 $\pm$ 0.028 & 0.496 $\pm$ 0.028 & 120.82
$\pm$ 0.46 & 341.61 $\pm$ \
0.51 & 115.3 $\pm$ 2.5 & 2000.95 $\pm$ 0.27 & 132. $\pm$ 11. & 14.7 \\
S2 & 0.123 $\pm$ 0.001 & 0.88 $\pm$ 0.003 & 135.25 $\pm$ 0.47 &
225.39 $\pm$ \
0.84 & 63.56 $\pm$ 0.84 & 2002.32 $\pm$ 0.01 & 15.8 $\pm$ 0.11 & 14. \\
S4 & 0.298 $\pm$ 0.019 & 0.406 $\pm$ 0.022 & 77.83 $\pm$ 0.32 &
258.11 $\pm$ \
0.3 & 316.4 $\pm$ 2.9 & 1974.4 $\pm$ 1. & 59.5 $\pm$ 2.6 & 14.4 \\
S5 & 0.25 $\pm$ 0.042 & 0.842 $\pm$ 0.017 & 143.7 $\pm$ 4.7 & 109.
$\pm$ 10. \
& 236.3 $\pm$ 8.2 & 1983.6 $\pm$ 2.5 & 45.7 $\pm$ 6.9 & 15.2 \\
S6 & 0.436 $\pm$ 0.153 & 0.886 $\pm$ 0.026 & 86.44 $\pm$ 0.59 &
83.46 $\pm$ \
0.69 & 129.5 $\pm$ 3.1 & 2063. $\pm$ 21. & 105. $\pm$ 34. & 15.4 \\
S8 & 0.411 $\pm$ 0.004 & 0.824 $\pm$ 0.014 & 74.01 $\pm$ 0.73 &
315.9 $\pm$ \
0.5 & 345.2 $\pm$ 1.1 & 1983.8 $\pm$ 0.4 & 96.1 $\pm$ 1.6 & 14.5 \\
S9 & 0.293 $\pm$ 0.052 & 0.825 $\pm$ 0.02 & 81. $\pm$ 0.7 & 147.58
$\pm$ 0.44 \
& 225.2 $\pm$ 2.3 & 1987.8 $\pm$ 2.1 & 58. $\pm$ 9.5 & 15.1 \\
S12 & 0.308 $\pm$ 0.008 & 0.9 $\pm$ 0.003 & 31.61 $\pm$ 0.76 &
240.4 $\pm$ \
4.6 & 308.8 $\pm$ 3.8 & 1995.63 $\pm$ 0.03 & 62.5 $\pm$ 2.3 & 15.5 \\
S13 & 0.297 $\pm$ 0.012 & 0.49 $\pm$ 0.023 & 25.5 $\pm$ 1.6 & 73.1
$\pm$ 4.1 \
& 248.2 $\pm$ 5.4 & 2004.9 $\pm$ 0.09 & 59.2 $\pm$ 3.8 & 15.8 \\
S14 & 0.256 $\pm$ 0.01 & 0.963 $\pm$ 0.006 & 99.4 $\pm$ 1. &
227.74 $\pm$ 0.7 \
& 339. $\pm$ 1.6 & 2000.07 $\pm$ 0.06 & 47.3 $\pm$ 2.9 & 15.7 \\
S17 & 0.311 $\pm$ 0.004 & 0.364 $\pm$ 0.015 & 96.44 $\pm$ 0.18 &
188.06 $\pm$ \
0.32 & 319.45 $\pm$ 3.2 & 1992. $\pm$ 0.3 & 63.2 $\pm$ 2. & 15.3 \\
S18 & 0.265 $\pm$ 0.08 & 0.759 $\pm$ 0.052 & 116. $\pm$ 2.7 &
215.2 $\pm$ 3.6 \
& 151.7 $\pm$ 2.9 & 1996. $\pm$ 0.9 & 50. $\pm$ 16. & 16.7 \\
S19 & 0.798 $\pm$ 0.064 & 0.844 $\pm$ 0.062 & 73.58 $\pm$ 0.61 &
342.9 $\pm$ \
1.2 & 153.3 $\pm$ 3. & 2005.1 $\pm$ 0.22 & 260. $\pm$ 31. & 16. \\
S21 & 0.213 $\pm$ 0.041 & 0.784 $\pm$ 0.028 & 54.8 $\pm$ 2.7 &
252.7 $\pm$ \
4.2 & 182.6 $\pm$ 8.2 & 2028.1 $\pm$ 5.5 & 35.8 $\pm$ 6.9 & 16.9 \\
S24 & 1.06 $\pm$ 0.178 & 0.933 $\pm$ 0.01 & 106.3 $\pm$ 0.93 & 4.2
$\pm$ 1.3 \
& 291.5 $\pm$ 1.5 & 2024.9 $\pm$ 5.5 & 398. $\pm$ 73. & 15.6 \\
S27 & 0.454 $\pm$ 0.078 & 0.952 $\pm$ 0.006 & 92.91 $\pm$ 0.73 &
191.9 $\pm$ \
0.92 & 308.2 $\pm$ 1.8 & 2059.7 $\pm$ 9.9 & 112. $\pm$ 18. & 15.6 \\
S29 & 0.397 $\pm$ 0.335 & 0.916 $\pm$ 0.048 & 122. $\pm$ 11. &
157.2 $\pm$ \
2.5 & 343.3 $\pm$ 5.7 & 2021. $\pm$ 18. & 91. $\pm$ 79. & 16.7 \\
S31 & 0.298 $\pm$ 0.044 & 0.934 $\pm$ 0.007 & 153.8 $\pm$ 5.8 &
103. $\pm$ \
11. & 314. $\pm$ 10. & 2013.8 $\pm$ 2.2 & 59.4 $\pm$ 9.2 & 15.7 \\
S33 & 0.41 $\pm$ 0.088 & 0.731 $\pm$ 0.039 & 42.9 $\pm$ 4.5 & 82.9
$\pm$ 5.9 \
& 328.1 $\pm$ 4.5 & 1967.9 $\pm$ 6.5 & 96. $\pm$ 21. & 16. \\
S38 & 0.139 $\pm$ 0.041 & 0.802 $\pm$ 0.041 & 166. $\pm$ 22. &
286. $\pm$ 68. \
& 203. $\pm$ 68. & 2003. $\pm$ 0.2 & 18.9 $\pm$ 5.8 & 17. \\
S66 & 1.21 $\pm$ 0.126 & 0.178 $\pm$ 0.039 & 135.4 $\pm$ 2.6 &
96.8 $\pm$ 2.9 \
& 106. $\pm$ 6.3 & 1782. $\pm$ 23. & 486. $\pm$ 41. & 14.8 \\
S67 & 1.1 $\pm$ 0.102 & 0.368 $\pm$ 0.041 & 139.9 $\pm$ 2.3 & 106.
$\pm$ 6.1 \
& 215.2 $\pm$ 4.8 & 1695. $\pm$ 16. & 419. $\pm$ 19. & 12.1 \\
S71 & 1.06 $\pm$ 0.765 & 0.844 $\pm$ 0.075 & 76.3 $\pm$ 3.6 & 34.6
$\pm$ 1.5 \
& 331.4 $\pm$ 7.1 & 1646. $\pm$ 251. & 399. $\pm$ 283. & 16.1 \\
S83 & 2.79 $\pm$ 0.234 & 0.657 $\pm$ 0.096 & 123.8 $\pm$ 1.3 &
73.6 $\pm$ 2.1 \
& 197.2 $\pm$ 3.5 & 2061. $\pm$ 25. & 1700. $\pm$ 205. & 13.6 \\
S87 & 1.26 $\pm$ 0.161 & 0.423 $\pm$ 0.036 & 142.7 $\pm$ 4.4 &
109.9 $\pm$ \
2.9 & 41.5 $\pm$ 3.7 & 1647. $\pm$ 38. & 516. $\pm$ 44. & 13.6 \\
S96 & 1.55 $\pm$ 0.209 & 0.131 $\pm$ 0.054 & 126.8 $\pm$ 2.4 &
115.78 $\pm$ \
1.93 & 231. $\pm$ 9. & 1624. $\pm$ 34. & 701. $\pm$ 81. & 10. \\
S97 & 2.19 $\pm$ 0.844 & 0.302 $\pm$ 0.308 & 114.6 $\pm$ 5. &
107.72 $\pm$ \
3.15 & 38. $\pm$ 52. & 2175. $\pm$ 88. & 1180. $\pm$ 688. & 10.3 \\
\tableline
\end{tabular}
\end{scriptsize}
\caption{Orbital Parameters of the S-stars examined in the paper:
$a$ is the semimajor axis, $e$ is the eccentricity, $i$ is the
inclination of the normal of the orbit with respect to the line of
sight, $\Omega$ is the position angle of the ascending node,
$\omega$ is the periapse anomaly with respect to the ascending
node, $t_{\mathrm{P}}$ is the epoch of last/next periapse, $T$ is
the orbital period, $K$ is the apparent magnitude in the $K$ band
(data taken from \citealt{Gillessen08}).} \label{tbl-1}
\end{table*}

\begin{table*}
\centering
\begin{scriptsize}
\begin{tabular}{lcccccc}
\tableline %
\tableline %
Star & $t_0[\mathrm{yr}]$ & $t_{1/2}[\mathrm{yr}]$ & $K_2$ &
$\theta_2[\mathrm{\mu as}]$ & $\gamma_0[^{\circ}]$ & $D_{\mathrm{LS},0}[AU]$ \\
\tableline
S1 & 2130.34 & 4.9 & 32.2 & 54 & 30.8 & 2195 \\
S2 & 2018.15 & 0.095 & 26.8 & 43 & 45.4 & 127 \\
S4 & 2048.94 & 3.9 & 28.7 & 112 & 12.2 & 2868 \\
S5 & 2028.8 & 2.7 & 32.9 & 32 & 83.6 & 456 \\
S6 & 2062.65 & 0.051 & 20.8 & 316 & 3.56 & 464 \\
S8 & 2082.7 & 1.6 & 28.3 & 88 & 16.2 & 1335 \\
S9 & 2041.39 & 1.9 & 27.1 & 144 & 9.06 & 1833 \\
S12 & 2058.38 & 1.4 & 32.6 & 32 & 88.5 & 319 \\
S13 & 2060.3 & 34 & 36.4 & 31 & 90.8 & 1571 \\
S14 & 2047.55 & 0.073 & 23.5 & 136 & 9.46 & 230 \\
S17 & 2008.24 & 2.1 & 26.9 & 196 & 6.44 & 2942 \\
S18 & 2045.15 & 0.72 & 30.9 & 61 & 26.2 & 670 \\
S19 & 2262.63 & 1.3 & 29.9 & 87 & 16.5 & 1369 \\
S21 & 2027.23 & 1.1 & 32.3 & 49 & 36.5 & 617 \\
S24 & 2047.95 & 44 & 33.1 & 81 & 17.9 & 6225 \\
S27 & 2062.08 & 0.55 & 22.4 & 399 & 2.92 & 1397 \\
S29 & 2021.73 & 0.98 & 31.6 & 51 & 34.1 & 572 \\
S31 & 2013.9 & 0.56 & 31.7 & 31 & 91.2 & 189 \\
S33 & 2067.95 & 9.1 & 34.9 & 39 & 53.7 & 1637 \\
S38 & 2021.8 & 0.71 & 33.5 & 32 & 88.9 & 242 \\
S66 & 2254.09 & 66 & 36.7 & 43 & 45.4 & 8328 \\
S67 & 2035.14 & 116 & 34.6 & 40 & 51.5 & 9053 \\
S71 & 2059.51 & 8.3 & 31.6 & 100 & 13.9 & 4073 \\
S83 & 3647.18 & 129 & 35.6 & 50 & 35. & 14435 \\
S87 & 2185.1 & 52 & 35.5 & 39 & 53.2 & 6508 \\
S96 & 2073.74 & 131 & 32.2 & 48 & 36.8 & 14049 \\
S97 & 2265.5 & 68 & 30.9 & 64 & 24.6 & 13897 \\
\tableline
\end{tabular}
\end{scriptsize} \caption{Summary of the main features of the
secondary images of the S-stars: $t_{0}$ is the epoch of the peak,
$t_{1/2}$ is the time spent by the secondary image at a luminosity
higher than half the maximum, $K_{2}$ is the peak $K$-band
magnitude for the secondary image, $\theta_{2}$ is the maximal
angular distance from the apparent shadow of the black hole,
$\gamma_{0}$ is the angle formed by the line connecting the
selected star with the central black hole and the optical axis at
the time of the peak, and $D_{\mathrm{LS},0}$ is the distance of
the S-star from the MBH at this time.} \label{tbl-2}
\end{table*}

\acknowledgments

We acknowledge support for this work by MIUR through PRIN 2006
Protocol 2006023491\_003, by research funds of Agenzia Spaziale
Italiana, by funds of Regione Campania, L.R. n.5/2002, year 2005
(run by Gaetano Scarpetta), and by research funds of Salerno
University.

\end{document}